\documentclass[aps,prstab,
reprint,
]{revtex4-2}

\usepackage[utf8]{inputenc} 

\usepackage{graphicx} 
\usepackage{dcolumn}
\usepackage{bm}

\usepackage{amsmath}
\usepackage[%
  colorlinks=true,
  urlcolor=blue,
  linkcolor=black,
  citecolor=blue
]{hyperref}

\draft 

\begin{document}


\title{Simulation study of betatron radiation in AWAKE Run 2 experiment} 



\author{Linbo Liang}
\email[]{linbo.liang@postgrad.manchester.ac.uk}
\affiliation{Department of Physics and Astronomy, University of Manchester, Manchester M13 9PL, United Kingdom}
\affiliation{Cockcroft Institute, Daresbury, Cheshire WA4 4AD, United Kingdom}

\author{Guoxing Xia}
\affiliation{Department of Physics and Astronomy, University of Manchester, Manchester M13 9PL, United Kingdom}
\affiliation{Cockcroft Institute, Daresbury, Cheshire WA4 4AD, United Kingdom}

\author{Hossein Saberi}
\affiliation{Department of Physics and Astronomy, University of Manchester, Manchester M13 9PL, United Kingdom}
\affiliation{Cockcroft Institute, Daresbury, Cheshire WA4 4AD, United Kingdom}

\author{John Patrick Farmer}
\affiliation{CERN, Geneva, Switzerland}
\affiliation{Max Planck Institute for Plasma Physics, Munich, Germany}

\author{Alexander Pukhov}
\affiliation{Heinrich-Heine-Universität Düsseldorf, Düsseldorf, Germany}


\date{\today}

\begin{abstract}
The spectroscopy of betatron radiation from the focusing plasma column can work as a powerful non-invasive beam diagnostic method for plasma wakefield acceleration experiments such as the AWAKE. In this paper, the effects of radial size mismatch and off-axis injection on the beam dynamics, as well as the spectral features of the betatron radiation emitted by the witness electron bunch in the quasi-linear proton-driven plasma wakefield are studied. 
It is shown that the evolution of the critical betatron photon energy and the overall photon angular distribution can effectively reveal the initial injection conditions of the witness electron bunch. The possibility of using this method for the diagnostics of the seed electron bunch in the proton self-modulation stage of AWAKE Run 2 is also discussed.
\end{abstract}


\maketitle 

\section{Introduction\label{sec:intro}}
Plasma wakefield acceleration has been proven to be a very attractive new acceleration concept due to not only its large acceleration gradients, but also its high focussing fields. 
In such intense fields, relativistic electrons can emit synchrotron-like broadband radiation when they oscillate transversely in the focusing plasma ion column~\cite{PhysRevE.65.056505, PhysRevLett.88.135004, doi:10.1063/1.1624605}, which is known as betatron radiation. 
With the advances of laser and plasma techniques, betatron radiation from plasma wakefield accelerators has become a valuable short-pulse broadband x-ray source for imaging with high contrast~\cite{RN907, RN908}.
In addition, it also works as an important beam diagnostic tool
~\cite{PhysRevLett.97.225002, PhysRevE.77.056402, PhysRevLett.111.235004, RN909, doi:10.1063/1.4773784}.

Conventional emittance diagnostic methods such as the pepper-pot~\cite{PhysRevSTAB.13.092803} and the quadrupole scan method~\cite{PhysRevSTAB.15.111302} are less effective when measuring the plasma accelerated electron bunches, which typically have few-GeV energy, few-fs duration, initially sub-$\mathrm{\mu m}$ radius, few-mrad divergence, and few-percent energy spread. 
Furthermore, the plasma-vacuum boundary can reshape the beam's transverse phase-space, thereby changing the downstream electron beam divergence, so the emittance measurement for particle beams inside the plasma is crucial~\cite{PhysRevSTAB.13.092803, PhysRevSTAB.15.111302}. 
These challenges can be circumvented with the betatron radiation diagnostics, which indirectly measure the accelerated electron bunch inside the plasma~\cite{PhysRevLett.97.225002, KOHLER2016265}. 

The Advanced Wakefield Experiment (AWAKE) is a proof-of-principle proton beam driven plasma wakefield experiment at CERN~\cite{AWAKENature}. 
The AWAKE Run 1 experiment (2016-2018) has achieved the seeded self-modulation (SSM) of the proton bunch and the acceleration of externally injected 18.5 MeV electrons to the 2 GeV~\cite{AWAKENature}. Following the success of Run 1, the AWAKE Run 2 experiment (2021-) aims to demonstrate the electron seeded proton self-modulation (eSSM), as well as to accelerate the externally injected electron bunch to a higher energy, e.g. $\sim$10 GeV, while controlling the beam emittance and energy spread~\cite{Muggli_2020}. The Run 2 setup will employ two cascaded 10-meter-long plasma cells with a narrow gap in between. Due to the aforementioned deficiencies of the conventional beam diagnostic methods as well as the challenging diagnostic environment along the beam propagation direction in AWAKE, a non-invasive beam diagnostic method for electron bunches in plasma is needed. 
A preliminary study~\cite{WILLIAMSON2020164076} has examined the possibility of using the betatron radiation to reconstruct the accelerated electron bunch's transverse profile. However, it has been shown that the evolving beam properties due to acceleration make it difficult to retrieve the beam profile and emittance via those methods for short-distance laser wakefield acceleration (LWFA) based experiments~\cite{PhysRevAccelBeams.20.012801, doi:10.1063/1.4998932}. 

In this paper, we systematically investigate the betatron radiation effect in AWAKE Run 2 by considering realistic Run 2 beam and plasma parameters. It is shown that possible non-ideal injection conditions of the witness electron beams, such as beam radius mismatch and transverse misalignment, can  significantly affect the witness beam dynamics and thus the spectral features of its betatron radiation. The radiation properties of seed electron bunch in the proton self-modulation stage is also studied.

This paper is organized as follows. The theoretical description of the betatron radiation in plasma ion column is presented in Section~\ref{theory}. The betatron radiation emitted by the mismatched as well as off-axis injected witness bunch in the acceleration stage is discussed in Section~\ref{witnessradiation}. The radiation from the low energy seed electron bunch in the self-modulation stage and the possibility for beam diagnostics is explored in Section~\ref{seedradiation}. Other conditions that affect the betatron radiation diagnostics are discussed in Section~\ref{discussection}.

\section{Betatron radiation in plasma ion column}
\label{theory}
The plasma ion column driven by a laser pulse or particle bunch is a place where both intense longitudinal acceleration force and transverse restoring force exist
~\cite{doi:10.1063/1.1624605}. The latter can lead to betatron oscillations of off-axis electrons with the fundamental frequency of $\omega_\beta=\omega_p/\sqrt{2\gamma}$ and the wavelength of $\lambda_\beta=2\pi c/\omega_\beta$. Here, $\gamma=E/m_ec^2+1$ is the Lorentz factor, $E$ is the beam energy, $c$ is the speed of light, $e$ and $m_e$ are the electron charge and mass, $\omega_p=\sqrt{n_0e^2/m_e\varepsilon_0}$ is the plasma wave frequency, $n_0$ is the unperturbed plasma electron density, and $\varepsilon_0$ is the vacuum permittivity. 

The strength of the betatron oscillations in plasma is usually characterized by the normalized betatron oscillation amplitude $K_\beta$, which is defined as~\cite{doi:10.1063/1.1624605}
\begin{equation}
\label{Kbeta}
K_\beta=\gamma k_\beta r_\beta=1.33\times10^{-10}\sqrt{\gamma n_e\mathrm{[cm^{-3}]}}r_\beta[\mathrm{\mu m}],
\end{equation}
where $k_\beta=k_p/\sqrt{2\gamma}$ is the betatron wave number for pure plasma ion background, $k_p=\omega_p/c$ is the plasma wave number, and $r_\beta$ is the betatron oscillation amplitude.
Depending on the strength of $K_\beta$, the betatron radiation is mainly categorised into two regimes. The limit of small amplitude near axis betatron oscillations with $K_\beta\ll1$ is known as the undulator regime. 
The undulator regime radiation is narrowly peaked at the fundamental mode ($n=1$) with wavelength of $\lambda=\lambda_\beta/2\gamma^2$~\cite{PhysRevE.65.056505}.  
On the other hand, if the betatron oscillation amplitude is large enough, i.e., $K_\beta\gg1$, radiations from different sections of the electron's trajectory will be emitted in different directions, contributing to a wide opening angle $\Psi=K_\beta/\gamma$ of the radiation cone in the direction perpendicular to the electron oscillation plane. The radiation frequency range also gets wide as high harmonics with finite bandwidth become significant. This regime is known as the wiggler regime and is valid for typical laser plasma wakefield accelerators (LWFA). 

Finite variations of the plasma wiggler parameter $K_\beta$ 
for different electrons in the bunch will broaden the bandwidth of each harmonic and lead to an overlap of different frequency spikes. So the betatron radiation spectrum in the plasma ion column appears as a quasi-continuous broadband spectrum, similar to the synchrotron radiation from a bending dipole magnet. This synchrotron-like spectrum has been demonstrated experimentally in laser-plasma experiments~\cite{PhysRevLett.93.135005, RN907, Fourmaux_2011}.

An asymptotic spectrum of the wiggler regime radiation of a single electron in the perpendicular direction is given by~\cite{PhysRevE.65.056505}
\begin{equation}
\label{Eq: asyspecdwdO}
S_{\gamma,r_\beta}(\omega,\Omega)
\sim\frac{\gamma^2\zeta^2}{1+\gamma^2\theta^2}\left[K_{2/3}^2(\zeta)+\frac{\gamma^2\theta^2}{1+\gamma^2\theta^2}K_{1/3}^2(\zeta)\right],
\end{equation}
where $\omega$ is the radiation frequency, and $\Omega$ represents the radiation direction and $\mathrm{d}\Omega=\sin\theta\mathrm{d}\theta\mathrm{d}\phi$. Here, $\theta$ is the deflection angle relative to the electron propagation direction and $\phi$ is the azimuthal angle in the vertical plane. 
$K_{\nu}(\zeta)$ is the modified Bessel function of the second kind and $\zeta=({\omega}/{\omega_c})\left(1+\gamma^2\theta\right)^{3/2}$. 
The critical photon frequency $\omega_c$ is defined as~\cite{PhysRevE.65.056505}
\begin{equation}
\label{Eq. criteng}
\omega_c=3\gamma^2\omega_\beta K_\beta\propto r_{\beta0}\gamma_0^{1/4}\gamma^{7/4},
\end{equation}
where $r_{\beta0}$ and $\gamma_0$ are the initial betatron oscillation amplitude and the Lorentz factor, respectively. The critical photon frequency is meaningful since half of the radiation power is emitted above/below the frequency of $\omega_c$. The radiation intensity falls exponentially and becomes negligible for frequencies beyond $\omega_c$. 

The normalized photon energy spectrum is obtained by integrating Eq.~\eqref{Eq: asyspecdwdO} over all spatial angles ($\theta, \phi$), such that 
\begin{equation}
\label{Eq: asypecdIdw}
S_{\gamma,r_\beta}(\omega)\sim(\omega/\omega_c)\int_{2\omega/\omega_c}^\infty K_{5/3}(\omega/\omega_c)\mathrm{d}(\omega/\omega_c).
\end{equation}
In plasma wakefield accelerators, the spatial scale of electron bunches is typically on the order of $\sim\mathrm{\mu m}$, which is much larger than the betatron radiation wavelength ($\sim$sub-nm). So the radiation spectrum of a bunch can be simplified as the incoherent summation of the single electron spectrum~\cite{PhysRevE.65.056505, PhysRevAccelBeams.20.012801}, which reads as
\begin{equation}
\label{eq: bunchrad}
S(\omega)=\int\mathrm{d}\Omega\int\mathrm{d}\gamma\int\mathrm{d}r_\beta\Gamma(\gamma)R(r_\beta)S_{\gamma,r_\beta}(\omega,\Omega),
\end{equation}
where $R(r_\beta)=r_\beta P(r_\beta)$ is the weighted radial distribution about $r_\beta$, $P(r_\beta)$ is the probability density function or the weight, $\Gamma(\gamma)$ is the electron beam energy spectrum and the theoretical single electron spectrum $S_{\gamma,r_\beta}(\omega, \Omega)$ should be calculated through its complete form~\cite{PhysRevAccelBeams.20.012801}. 

A beam profile and emittance recovery method based on Eq.~\eqref{eq: bunchrad} has been proposed~\cite{PhysRevAccelBeams.20.012801, doi:10.1063/1.4998932}, which solves the inverse problem of Eq.~\eqref{eq: bunchrad} to get the beam profile $P(r)$ if the electron beam energy spectrum $\Gamma(\gamma)$ is measured. Additionally, the divergence term $\Theta(\theta_d)$ can be retrieved from the correlation between $\theta_d$ and $r_\beta$. However, as aforementioned, the evolution of the electron beam energy and its transverse profile during the acceleration may prohibit us to reconstruct the beam profile with this simple analytical model if the measured radiation is integrated over multiple betatron periods as in the case of AWAKE~\cite{WILLIAMSON2020164076}. A spectrometer that can resolve the single period emission (sub-fs) or a model that considers the electron beam evolution is required.

\section{Radiation from the witness electron bunch}
\label{witnessradiation}

\begin{figure}
\centering
\includegraphics*[width=\columnwidth]{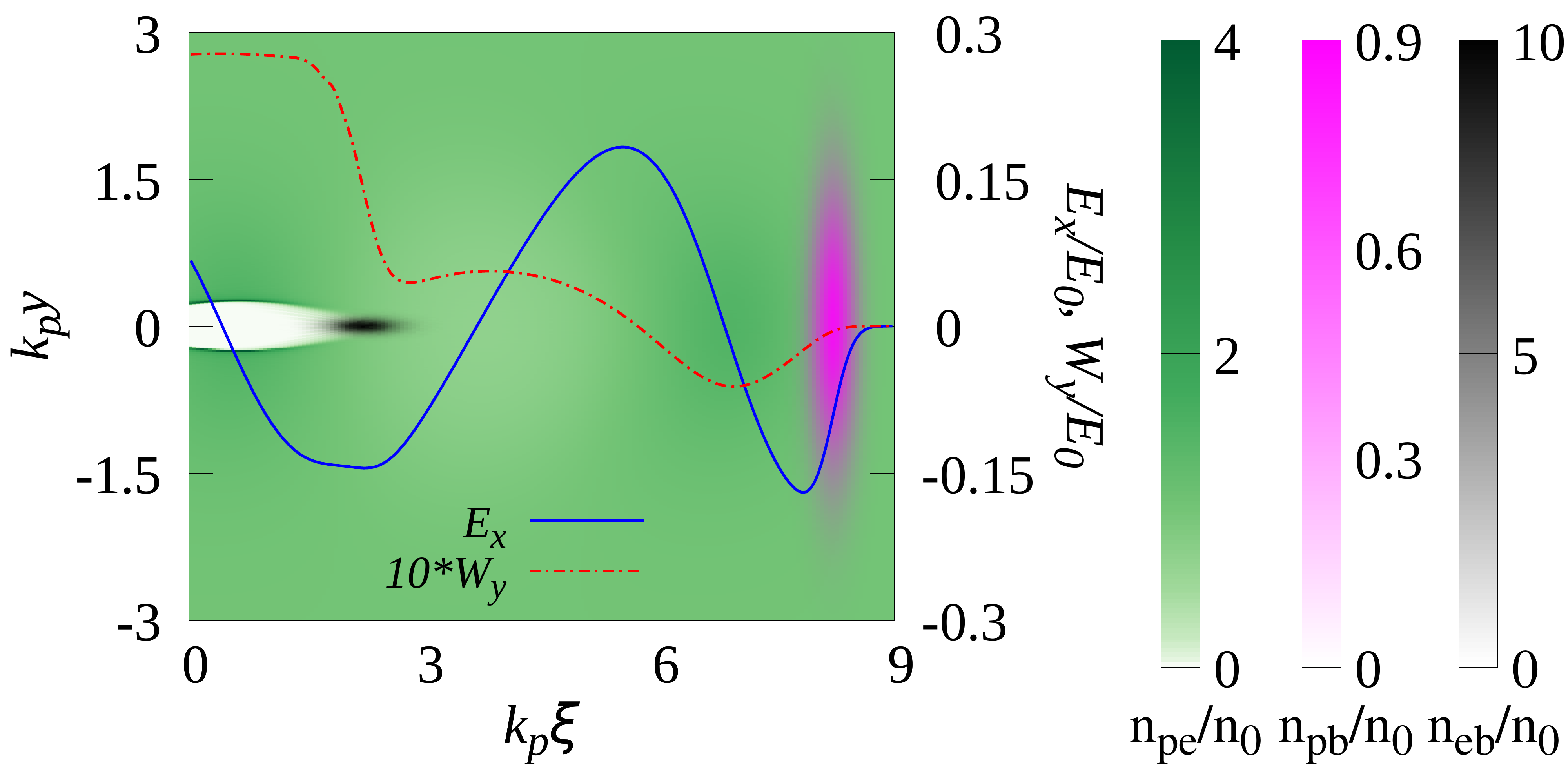}
\caption{The simplified schematic of AWAKE Run 2 acceleration stage. The densities of plasma electrons ($n_{pe}$, green), the proton driver bunch ($n_{pb}$, magenta) and the electron witness bunch ($n_{eb}$, black) are shown in contour plots. Particle beams propagate from left to right.  $\xi=x-ct$ is the longitudinal coordinate in the co-moving frame. The blue solid line represents the loaded longitudinal wakefield $E_x$, and the red dashed line is the transverse wakefield $W_y=E_y-cB_z$ at a transverse position of one $\sigma_{ic}$ from the longitudinal axis $x$. $\sigma_{ic}$ is the matched RMS beam radius of the electron witness bunch. 
\label{fig: toymodel}
} 
\end{figure}

\begin{table}
\caption{Baseline parameters for electrons acceleration simulation.}
\label{AWAKEpara}
\begin{center}
\begin{tabular}{llll}
\hline
Parameters&Symbol& Value& Unit\\
\hline
\textbf{Rubidium Plasma}&&&\\
Density &$n_0$&$7\times10^{14}$&cm$^{-3}$\\\\
\textbf{Proton Driver Bunch}&&&\\
Energy & $E_{p0}$           		 &  400      & GeV\\ 
Lorentz Factor &$\gamma_{p0}$ &426.31&\\
Charge &$Q_p$           	 	&  2.34  	& nC\\
RMS Bunch Length & $\sigma_{\xi p}$ & 40 & $\mathrm{\mu m}$\\
RMS Bunch Radius & $\sigma_{rp}$ & 200 & $\mathrm{\mu m}$\\\\
\textbf{Electron Witness Bunch}&&&\\
Energy & $E_{e0}$             		 & 150     & MeV\\ 
Lorentz Factor &$\gamma_{e0}$ &294.54&\\
Energy Spread &$\Delta\gamma/\gamma_{e0}$& 0.1\%&\\
Charge &$Q_e$           	 	&  120  	& pC\\
RMS Bunch Length& $\sigma_{\xi e}$ & 60 & $\mathrm{\mu m}$\\
Normalized Emittance & $\varepsilon_{n0}$ & 6.84 &$\mathrm{\mu m}$\\
\hline
\end{tabular}
\end{center}
\end{table}

For the AWAKE Run 2, the self-modulated proton beam from the first plasma cell is injected into the second plasma cell to drive the acceleration wakefield. The witness electron beam with an energy of 150 MeV ($\gamma_{e0}=294.541$) is injected in the wakefield right after proton microbunches. For simplicity, the acceleration scheme is described by a toy model that was first introduced by V. Olsen \textit{et al.}~\cite{Olsen2018}. The model consists of a single, highly-rigid proton bunch as the driver and a witness electron bunch trailing behind, as shown in Fig.~\ref{fig: toymodel}. The driver particle mass is magnified by $10^{10}$ times on the base of the real proton mass in simulation. Since such a dummy proton bunch is highly rigid, the proton-driven wakefield also remains static in its own frame during the propagation except for the dephasing with respect to the witness bunch. This model allows to focus on the acceleration physics of the witness electron beam only. 

Necessary particle beams and plasma parameters for simulations are presented in Table~\ref{AWAKEpara}. 
Parameters of the non-evolving driver bunch are set to simulate the quasi-linear wakefield excited by the self-modulated SPS proton bunch train~\cite{Olsen2018}. 
The waist of the witness bunch is assumed to match to the pure plasma ion background at the entrance of the plasma column to prevent beam envelope or root-mean-square (RMS) beam radius oscillations. The mismatch of the beam radius can lead to a significant beam emittance growth~\cite{Olsen2018, injectiontolerance}. The matched radial beam size for a Gaussian beam in the plasma ion column is defined by~\cite{Olsen2018, doi:10.1098/rsta.2018.0181}:
\begin{equation}
\sigma_{ic}=\left(\frac{2\epsilon_{n0}^2}{\gamma_{e0}k_p^2}\right)^{1/4},
\label{matchcondition}
\end{equation}
where $\epsilon_{n0}=\beta_{e0}\gamma_{e0}\epsilon_0$ is the normalized RMS beam emittance, $\epsilon_0$ is the initial geometric emittance, $\gamma_{e0}$ and $\beta_{e0}=\sqrt{1-1/\gamma_{e0}^2}$ are the initial mean Lorentz factor and relativistic velocity of the witness beam. For the baseline witness beam parameters in Table~\ref{AWAKEpara}, Eq.~(\ref{matchcondition}) gives a matched beam size of $\sigma_{ic}=10.64$ $\mathrm{\mu m}$, resulting in a normalized peak bunch density of $n_{e0}/n_0\approx10$ for a Gaussian profile witness bunch. 

Since the baseline witness bunch is dense enough, it is able to further blow out the plasma electrons, forming a plasma bubble with linear radial focusing force after the bulk of itself. The longitudinal wakefield excited by witness also loads upon the proton-driven wakefield, resulting in a relatively uniform net accelerating gradient along the bunch. The initial delay between the two bunches is fixed to $k_p\Delta\xi=6$ for all cases presented in this work.

Numerical simulations of the witness bunch acceleration are carried out with the three-dimensional (3D) quasi-static particle-in-cell (PIC) code QV3D~\cite{PIC_Pukhov}, which is developed on the VLPL platform~\cite{pukhov_1999}. Since quasi-static PIC codes cannot explicitly model the radiation, 
QV3D calculates the radiation spectrum with the aforementioned analytical model per marcro-particle, per timestep. 
The simulation window co-moving with the particle bunches with a speed-of-light has the dimension of $(9\times6\times6)k_p^{-1}$ and resolution $(0.05\times0.01\times0.01)k_p^{-1}$ in directions of $(x,y,z)$, where $x$ is the longitudinal direction, $y$ and $z$ are transverse directions. The simulation timestep is chosen as $5\omega_p^{-1}$, which is enough to resolve the envelope evolution of the witness bunch. The number of macro-particles per cell is 4 for the plasma with fixed ion background and 1 for the non-evolving proton driver. The witness beam is simulated with $10^6$ equally-weighted macro particles.

\subsection{Effect of mismatched beam radius\label{sec:mismatch}}
In experiment, it is always difficult to precisely match the electron beam to the pure ion channel due to various errors in the beam transportation. Additionally, the plasma bubble with longitudinally uniform focusing strength covers only the rear part of the witness beam, while at its head the plasma focusing strength varies with the sinusoidal quasilinear plasma wave, thereby the matching conditions are not entirely identical for different slices along the witness beam. In other words, for the same initial beam radius of $\sigma_{ic}$ in all slices, the beam matching is only achieved inside the plasma bubble, whereas the beam head is mismatched with the quasilinear wakefield with weak focusing strength. With different extent of mismatch along the beam, the witness beam undergoes a fast and intense envelope expansion and oscillation after being injected into the plasma, especially at the bunch head. This leads to a rapid emittance growth at the initial period of beam propagation until full phase-mixing~\cite{liang:ipac2021-wepab175, injectiontolerance}. 
Nevertheless, Ref.~\cite{injectiontolerance} shows that the mismatch of whole bunch's RMS radius doesn't necessarily leads to the further degradation of the beam quality. There is actually a wide mismatch tolerance depending on the beam charge and length. The optimal initial bunch radius for minimal emittance growth is found to be larger than the matched bunch radius in the pure plasma ion column.

\begin{figure}
\centering
\includegraphics*[width=0.8\columnwidth]{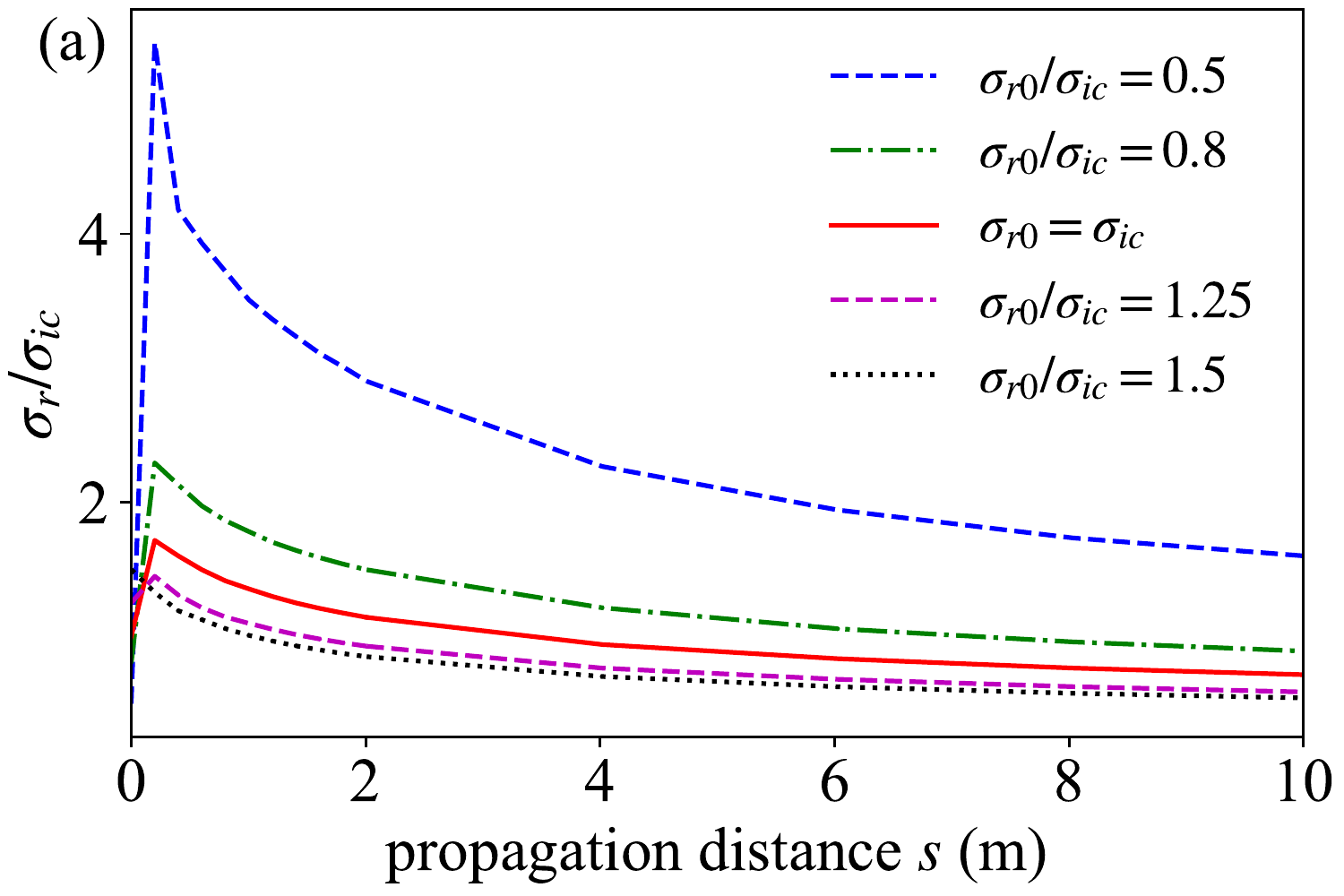}
\centering
\includegraphics*[width=0.8\columnwidth]{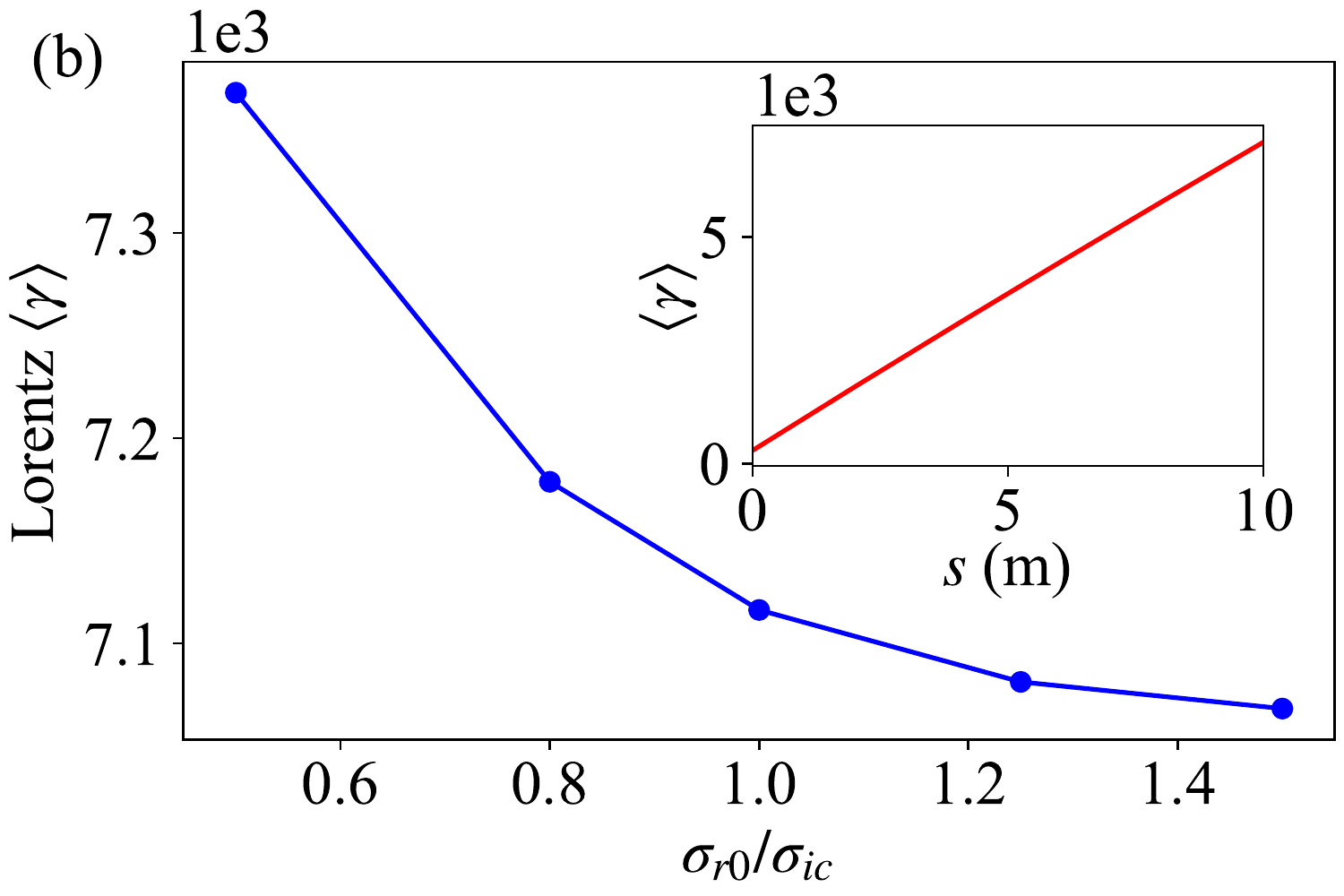}
\caption{(a) The RMS beam radius $\sigma_r$ versus the acceleration distance $s$. (b) Dependency of the average Lorentz factor $\left<\gamma\right>$ on the normalized initial beam radius $\sigma_{r0}/\sigma_{ic}$, measured after 10 m propagation in plasma. The insert shows the evolution of $\left<\gamma\right>$ for the matched case during the propagation.
\label{mismatch_radius}
}
\end{figure}

Figure~\ref{mismatch_radius}(a) shows the evolution trends of the RMS beam radius during the acceleration process for five cases with different initial beam radii. 
For cases with initial radius smaller than the matched one, the bubble formation is much quicker at the beginning. However, the witness bunch is over dense so the emittance pressure or emittance induced defocusing force exceeds the plasma focusing force both inside and outside the bubble. This leads to the rapid expansion of the beam envelope along the whole bunch in the first few tens of centimetres. The transverse expansion of bunch size is most pronounced at the bunch head, because as the transverse focusing force due to the quasi-linear wake is much weaker than in the bubble. In such cases, the RMS beam radius of the whole bunch after 10 m is larger than the ``matched'' case. The reduction of bunch radius with respect to the maximum value is due to the adiabatic damping effect, with the scaling law of $\sigma_r\propto\gamma^{-1/4}$. For cases with initial RMS radius larger than the matched value, the aforementioned physical process is reversed for the rear part of the beam inside the plasma bubble. However, for its head part, there is an ``optimal'' or quasi-matched initial RMS radius in the range of $1.25\leq\sigma_{r0}/\sigma_{ic}\leq 1.5$. Below this value, the radius of the head part first sees an expansion before the acceleration-leaded damping, like cases with $\sigma_{r0}<\sigma_{ic}$. And when the initial beam radius exceeds this ``optimal'' value, the emittance pressure is lower than the plasma focusing force along the bunch, so the entire bunch remains focused during whole process of acceleration. 

In Fig.~\ref{mismatch_radius}(a), a larger initial bunch radius results in a smaller value of the RMS beam radius after the first half meter, thus a larger average bunch density is obtained. This causes the overloading of the proton-driven wakefield, which then leads to a lower average accelerating gradient than cases with lower initial bunch radii. As a consequence, the final energy gain is lower for cases with the larger initial bunch radii, as shown in Fig.~\ref{mismatch_radius}(b). However, since the variation of the quasi-uniform accelerating gradient due to the mismatch and adiabatic damping effects is relatively small, the difference of the final energy gain for different mismatched cases is also small, and the average energy gain increases quasi-linearly during the acceleration.

\begin{figure}[h]
\centering
\includegraphics*[width=0.8\columnwidth]{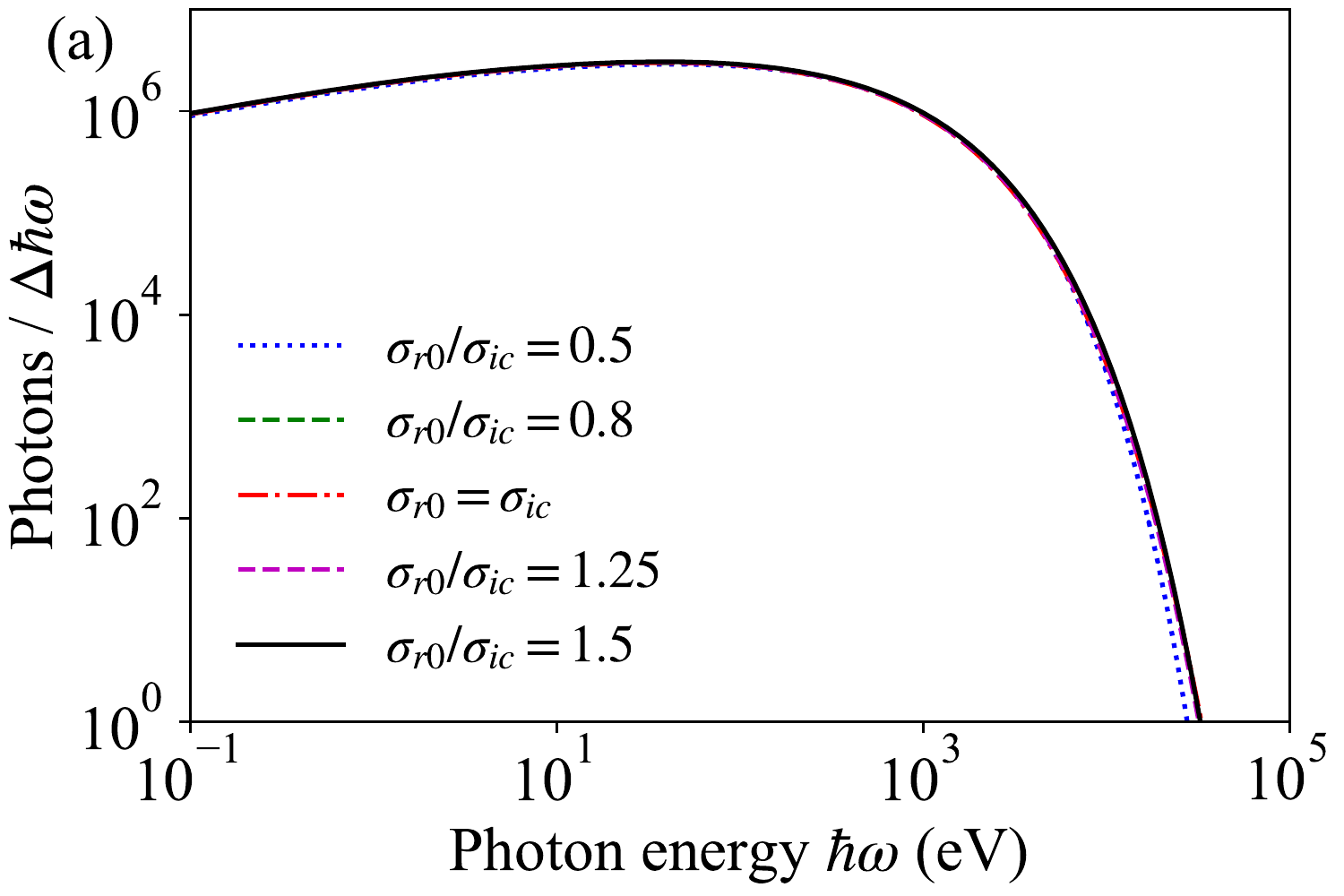}
\includegraphics*[width=0.8\columnwidth]{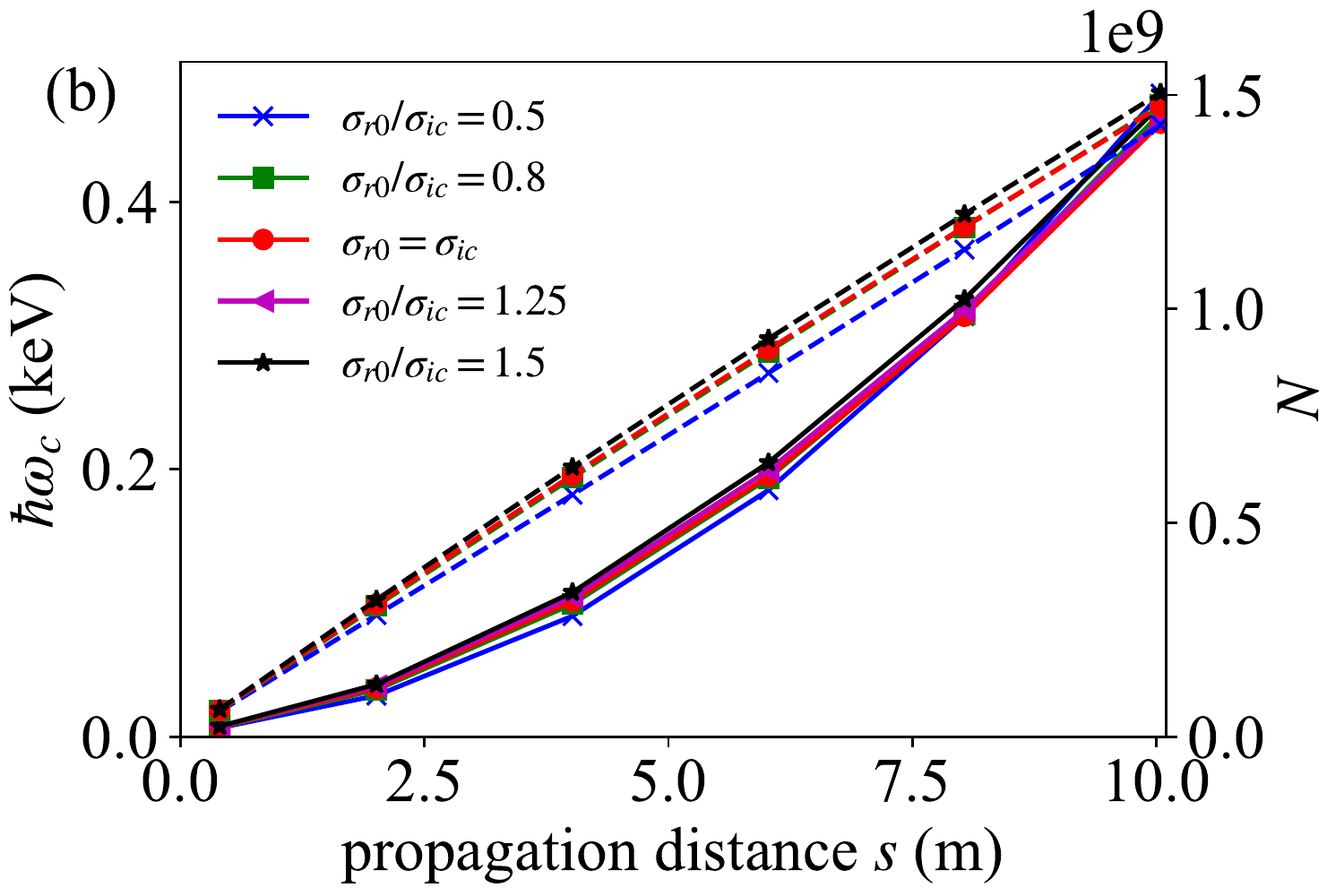}
\includegraphics*[width=0.8\columnwidth]{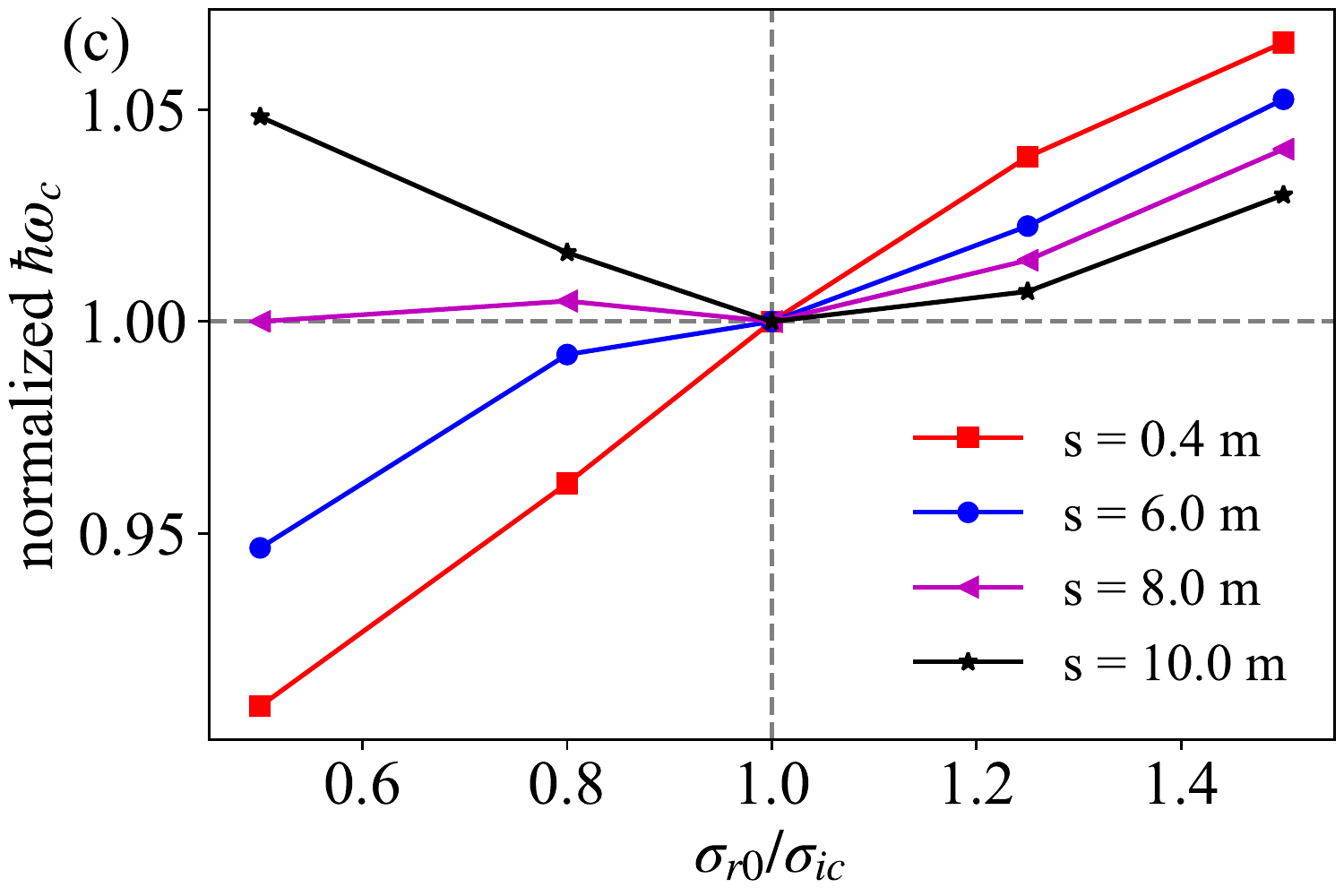}
\caption{Simulation obtained betatron radiation spectrum. (a) Photon energy spectrum measured at $s=10$ m. $\Delta\hbar\omega$ is $10^{-3}$ of the photon energy measuring range, given by the horizontal axis. (b) Evolution of the critical photon energy $\hbar\omega_c$ (solid lines) and total number of emitted photons $N$ (dashed lines) along the acceleration distance $s$. (c) Normalized critical photon energy $\hbar\omega_c$ (with respect to the value of matched case) vs. the normalized initial beam radius $\sigma_{r0}/\sigma_{ic}$.
\label{mismatch_spectrum}
}
\end{figure}
Figure~\ref{mismatch_spectrum}(a) shows the betatron photon energy spectra emitted by the witness electron bunches with different initial bunch radii. However, the difference is almost negligible for those considered cases. This is possibly due to that these betatron spectra are results of the emission integrated over multiple betatron oscillations up to the diagnostic point at $s$ = 10 m in our simulations. 
In order to quantitatively characterize the betatron photon spectra shown in Fig.~\ref{mismatch_spectrum}(a), the mean photon energy $\hbar\left<\omega\right>$ can be calculated. Additionally, we can estimate the critical photon energy via the relation~\cite{PhysRevE.65.056505}
\begin{equation}
\label{eq:critavg}
\hbar\omega_c=\hbar\left<\omega\right>/\left(4/15\sqrt{3}\right).
\end{equation}
Such a relation between $\hbar\left<\omega\right>$ and $\hbar\omega_c$ is originally derived from the single electron betatron radiation theory, so it is essentially not suitable for characterizing the integrated spectrum that consists a series of single electron emissions with various theoretical critical photon energies. 
Nevertheless, Fig.~\ref{mismatch_spectrum}(b) shows that the evolution trends of the critical photon energies of different beam cases generally follow the exponential growth relation given by Eq.~\eqref{Eq. criteng}, i.e., $\omega_c\propto\gamma^{7/4}$. This suggests that the single electron radiation theory is generally valid for the integrated bunch radiation and $\omega_c$ can still work as a useful figure of merit to evaluate the integrated betatron emission in our simulations. 

Fig.~\ref{mismatch_spectrum}(b) also shows the total number of photons $N$ emitted during the acceleration, which is almost linearly increasing during the beam propagation. Early researches have suggested that the total photon emission of a single electron scales with the the number of oscillations $N_\beta$ and its oscillation amplitude $K_\beta$, i.e., $N\propto N_\beta K_\beta$ for the wiggler regime with $K_\beta\gg 1$ and $N\propto N_\beta K_\beta^2$ for the undulator regime with $K_\beta<1$~\cite{PhysRevE.65.056505, RevModPhys.85.1}. In the case with constant energy gain during the acceleration, i.e., $\gamma\propto L$ with $L$ being the acceleration distance, and the electron oscillating in the pure ion column, the scaling laws of betatron emission can be further reduced as $N\propto r_{\beta0}L^{3/4}$ for $K_\beta\gg 1$ and $N\propto r_{\beta0}^2L$ for $K_\beta<1$. For witness bunches with the baseline parameters, the majority of witness electrons within one $\sigma_{r}$ oscillate with $K_\beta\leq1$ at the beginning and then rise up to $K_\beta\approx5$ at $s$ = 10 m for the case of $\sigma_{r0}/\sigma_{ic}=0.5$. Thus, the total number of photons emitted by the witness bunches is expected to linearly increase with the propagation distance.

In order to illustrate the relative change of the betatron emission for different cases studied above, lineouts showing the dependencies of the critical photon energy on the initial bunch radius are also plotted in Fig.~\ref{mismatch_spectrum}(c), where the critical photon energies are normalized by the value of the matched case. One can see that at an early moment, e.g. $s$ = 0.4 m, the critical photon energy is nearly proportional to the initial beam radius due to the relation $\omega_c\propto r_{\beta_0}$. During the acceleration, the relative difference of $\hbar\omega_c$ for different bunch cases gets smaller due to the increased contribution of the high energy photon emission close to the plasma exit. And one can also notice that the critical photon energy of cases with smaller lower bunch radius exceeds that of the matched case after $s$ = 8 m. This is likely due to the over expansion of bunch head for witness beams with small initial radius ($\sigma_{r0}<\sigma_{ic}$), which results in the betatron oscillations with large amplitude for these cases.

\begin{figure}
\centering
\includegraphics*[width=0.8\columnwidth]{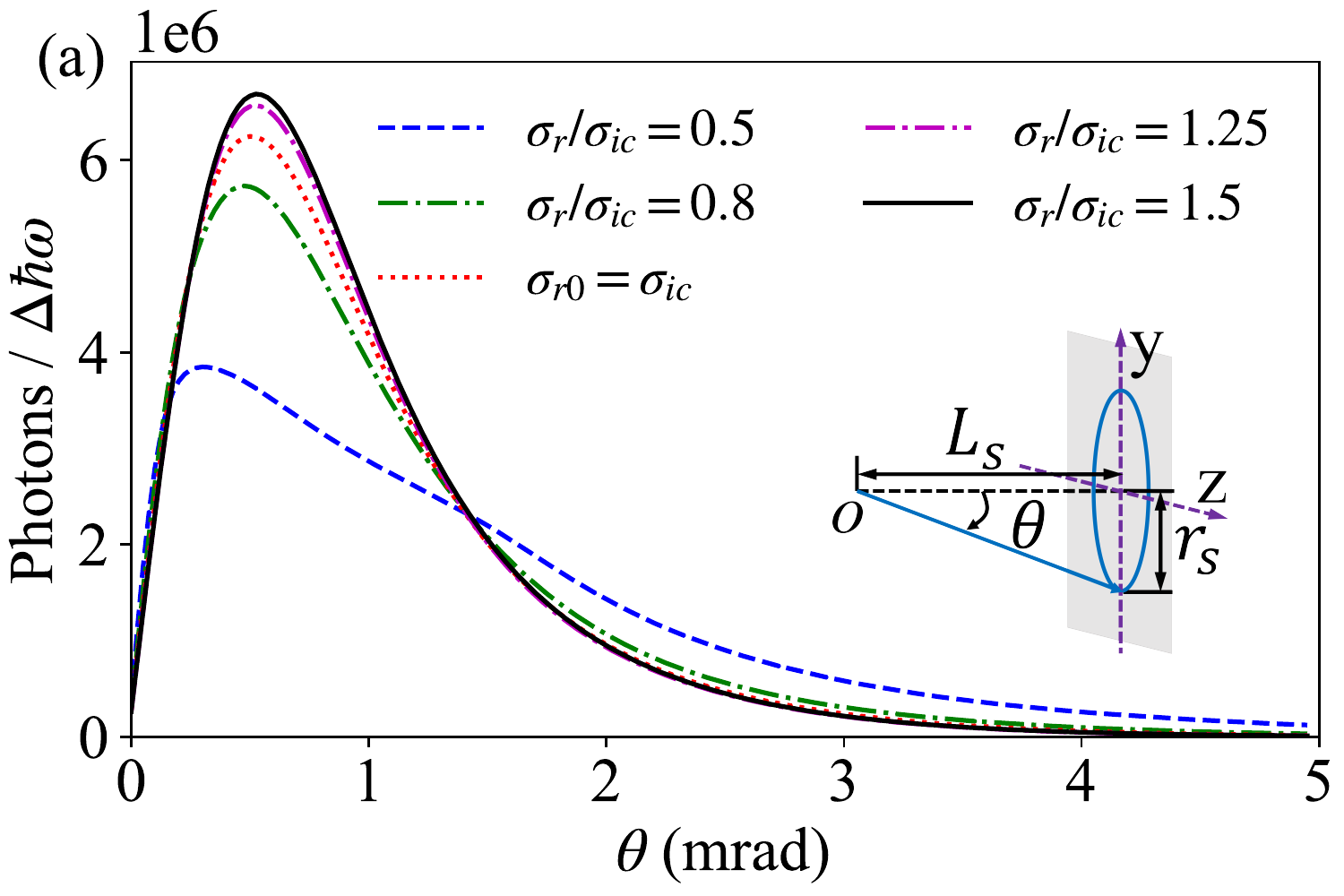}
\includegraphics*[width=0.8\columnwidth]{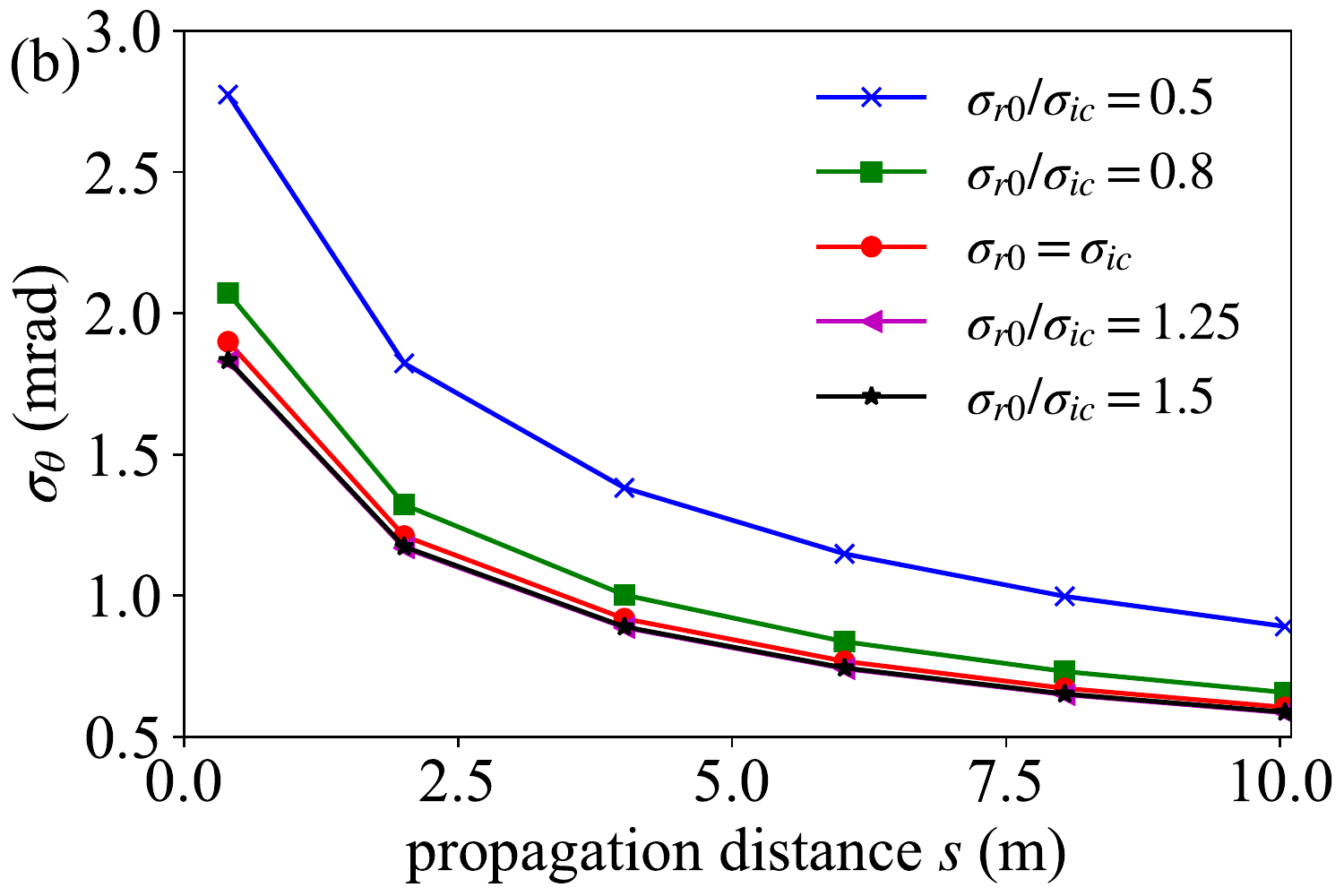}
\caption{(a) Photon angular distribution w.r.t the axial angle $\theta$ measured at $s=2$ m. (b) Evolution of the RMS value of $\theta$ versus the acceleration distance $s$.
\label{mismatch_angdist}
}
\end{figure}

For the purpose of beam profile and emittance reconstruction, the complete spatial distribution of a particle beam is needed~\cite{doi:10.1063/1.4998932}. So it is interesting to look at the spatial distribution of betatron photons. As the radiation pattern on the screen is axisymmetric for an axisymmetric electron bunch that is injected on the wakefield axis~\cite{doi:10.1063/1.1624605}, here we only look at the angular photon distribution with respect to the axial angle $\theta$. The angle $\theta$ represents the ratio between the radial position $r_s$ of betatron photons on a virtual screen and the distance $L_s$ between the screen and the plasma entrance, i.e., $\theta\approx r_{s}/L_s$, as illustrated in Fig.~\ref{mismatch_angdist}(a). 
Fig.~\ref{mismatch_angdist}(a) shows that the betatron photons emitted by the baseline witness electron bunches are confined within a narrow axial angle, with the RMS value of $\sigma_\theta=\left(\frac{1}{N}\sum_i\theta_i^2\right)^{1/2}<2$ mrad for photons measured at $s$ = 2 m. Although such angular photon distributions measured at the diagnostic point in our simulations are the integrated results, the photon angular distribution shape still have good agreement with the transverse distributions of the witness beams. 
Furthermore, we can see the evolution trend of $\sigma_\theta$ shown in Fig.~\ref{mismatch_angdist} (b) is similar to that of the RMS beam radius shown in Fig.~\ref{mismatch_radius}, although strictly speaking the typical axial angle of radiation scales to $K_\beta/\gamma$~\cite{PhysRevE.65.056505}, while the RMS radial beam size evolves with $1/\gamma^4$ in the plasma ion column.

\subsection{Effect of off-axis electron injection}
In section~\ref{sec:mismatch}, we assume the witness electron bunch is injected on the central axis of the proton-driven wakefield for ideal acceleration. The on-axis injected electron bunch doesn't suffer the possible transverse instabilities, e.g., hosing~\cite{PhysRevLett.99.255001}, which can be induced by the off-axis or oblique injection. This instability can cause a large increase of the beam emittance and even lead to the beam breakup if the instability gets strong enough. As the betatron oscillation initially depends on the transverse position of electrons at the injection point, it can be expected that the off-axis injection leads to stronger betatron emission with higher photon energy. The polarization of betatron radiation can also be enhanced in the preferred oscillation direction, i.e., the offset direction of the bunch~\cite{doi:10.1063/1.1624605}. This is due to the fact that transverse focusing force in the axisymmetric wakefields is pointing towards the axis of the wake structure. 

\begin{figure}
\centering
\includegraphics*[width=0.8\columnwidth]{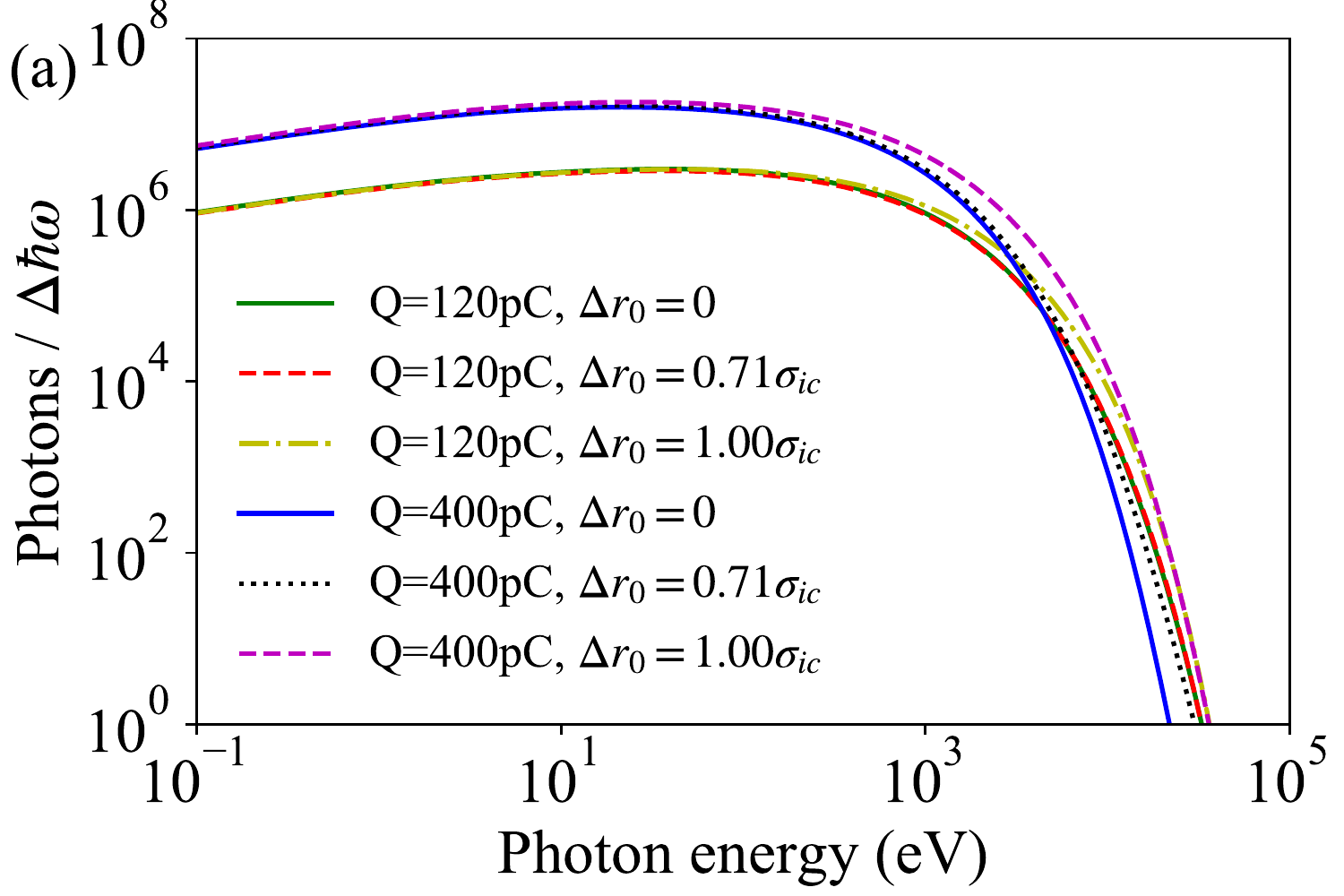}
\includegraphics*[width=0.8\columnwidth]{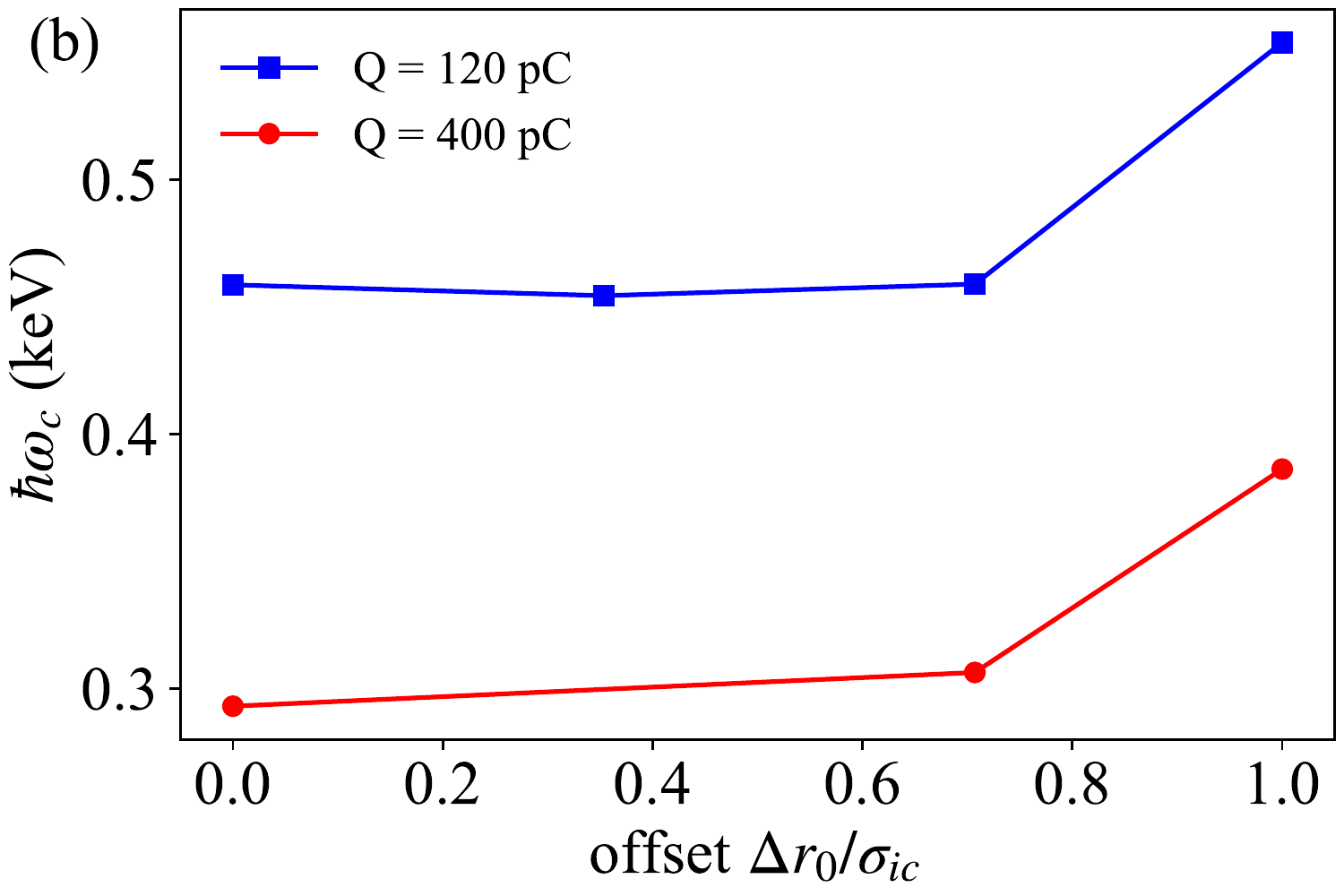}
\caption{(a) Betatron photon energy spectrum of different injection offsets at $s$ = 10 m. for Both the baseline witness beams and higher charge (400 pC) beams are considered. (b) The corresponding critical photon energies versus the beam offsets. 
\label{offset_spectrum}
}
\end{figure}

Here we look at several cases where the baseline witness bunches have minor offsets from the axis: $\Delta r_0/\sigma_{ic}=0$, 0.35, 0.71, and 1, where $\Delta r_0=\sqrt{\Delta y_0^2+\Delta z_0^2}$ is the combined initial offset of the transverse beam centroid, $y_0$ and $z_0$ are the initial offset in the $y$- and $z$-direction, respectively. For the 2nd and 3rd cases, the beam offsets are only in the $y$-direction, while for the last case, equal offsets of $0.71\sigma_{ic}$ present in both the $y$- and $z$-direction, corresponding to a combined offset of $\Delta r_0=\sigma_{ic}$ along the azimuthal angle $\phi=\pi/4$ in the $y-z$ plane.
These minor offsets allow to investigate the betatron radiation from the witness bunch without significant charge loss due to transverse instabilities. Additionally, the witness bunches with higher charge, e.g., 400 pC, are also studied. Higher charge is found to be able to compensate the beam emittance growth induced by the minor offset at the injection point, as the bubble formation is much quicker for the high charge bunch~\cite{injectiontolerance}. 

One can see in Fig.~\ref{offset_spectrum} that for the baseline witness bunches, the case with radial offset larger than $0.71\sigma_{ic}$ indeed radiates with higher critical photon energy than the on-axis injection case. For cases with offset smaller than that value, the amplification in emitted photon energy is not so significant. The trend can also be found in the radiation of the 400-pC witness bunches. This may suggest a safe range for driver-witness misalignment at the injection point. 
In Fig.~\ref{offset_spectrum}(a), one can also see that the 400-pC bunches can produce more betatron radiation than the baseline cases, which is mainly due to more charge engaging in the radiation process. Additionally, Fig.~\ref{offset_spectrum}(b) shows that the critical photon energies of cases with 400-pC charge are generally lower than the baseline witness bunches, which is essentially due to lower average accelerating wakefields for these 400-pC bunches after beam loading.

\begin{figure}
\centering
\includegraphics*[width=0.8\columnwidth]{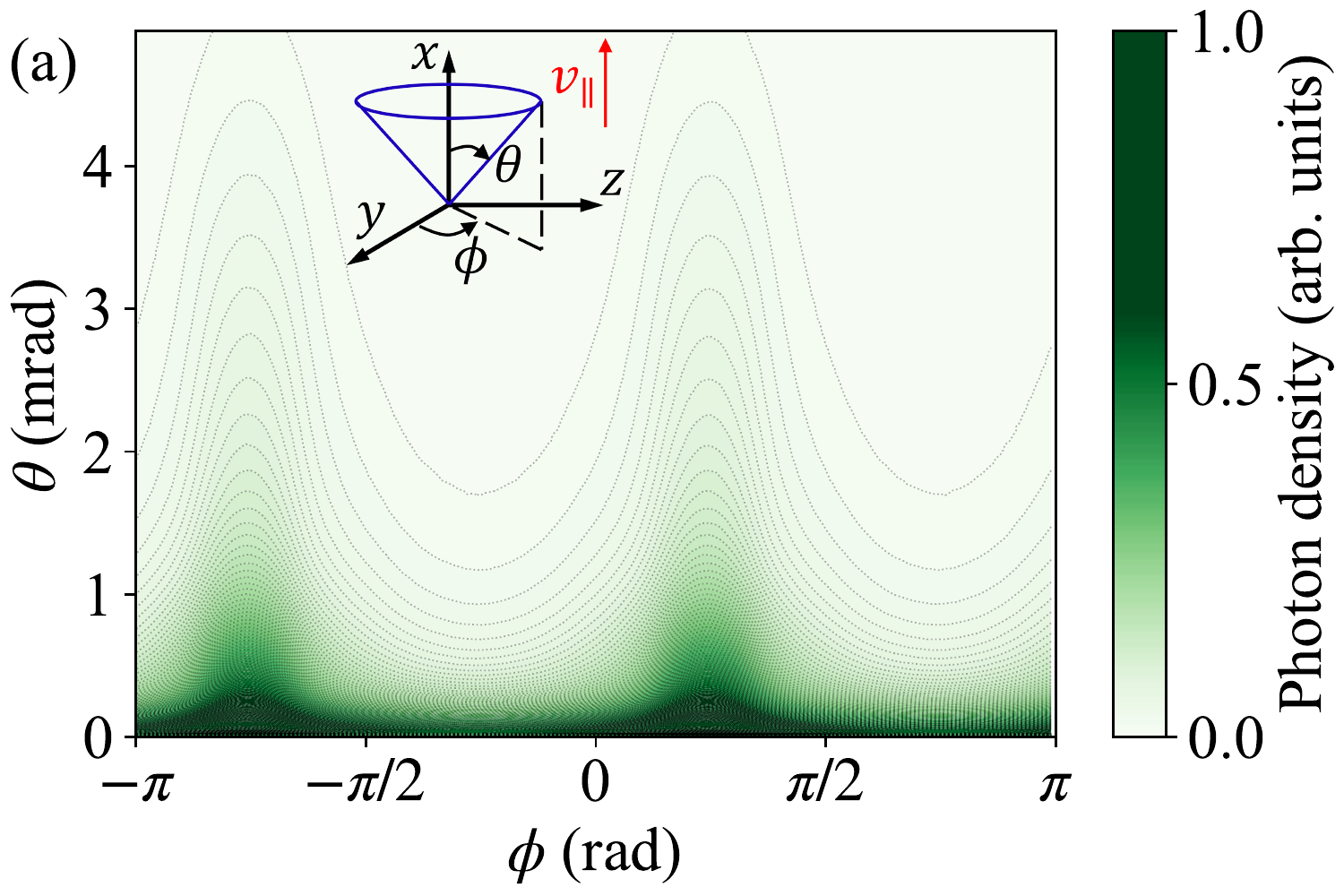}
\includegraphics*[width=0.8\columnwidth]{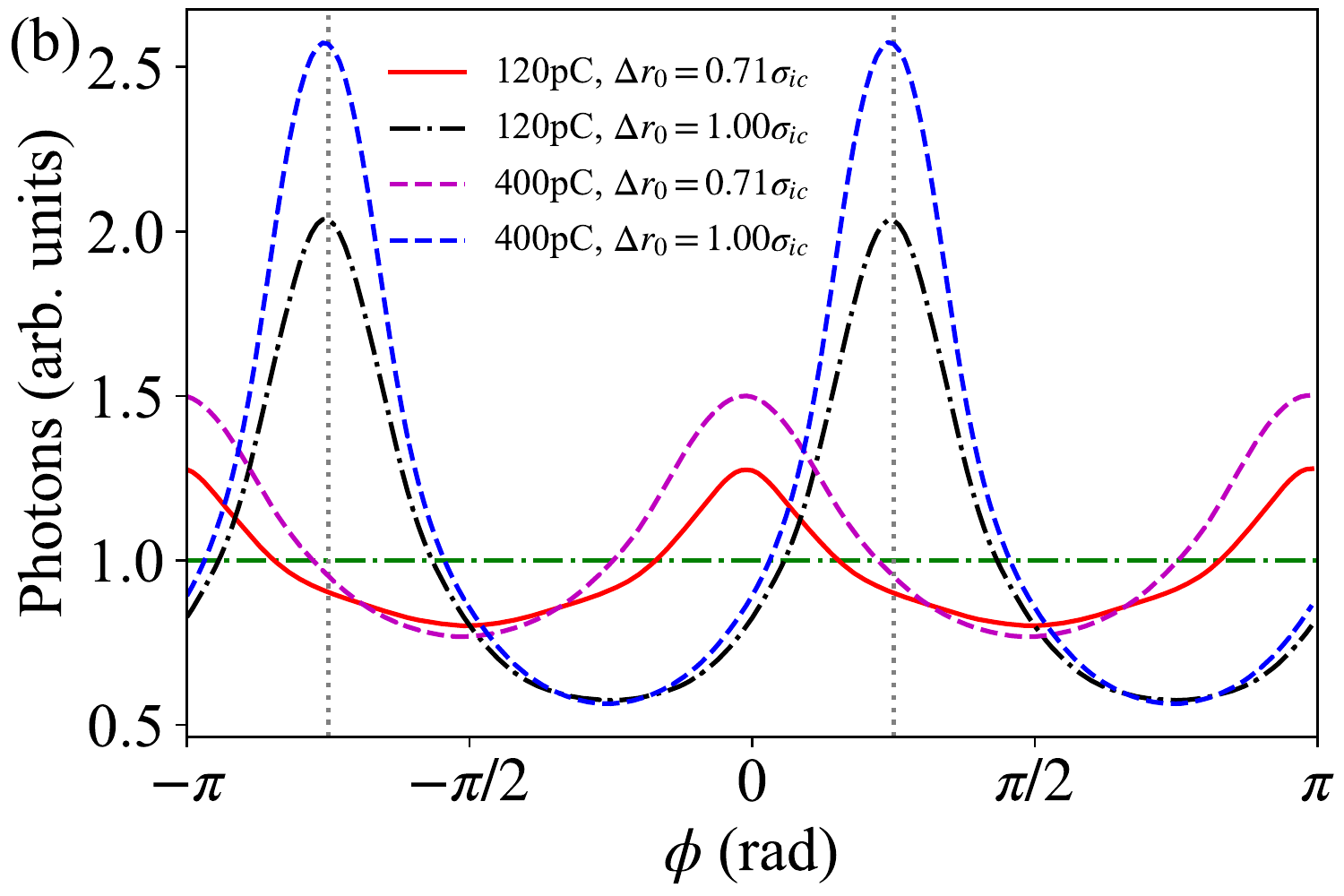}
\caption{Photon angular distribution for witness beam injection with transverse offsets. (a) 2D photon angular distribution of the baseline witness bunch with offsets of $0.71\sigma_{ic}$ in both the $y$- and $z$- direction. The inset shows the spherical coordinates used to represent the photon spatial distribution. (b) The 1D-dependencies of photon numbers on the azimuthal angle $\phi$. $\phi=0$ and $\pm\pi$ represents the $\pm y$ directions, respectively, and $\pm \pi/2$ are the $\pm z$ directions. The photon densities are normalized by the value of the onaxis injection cases for both the 120 pC and 400 pC. The azimuthal distribution of the onaxis injection cases is represented by the horizontal green dash-dotted line.
\label{offset_phi}
}
\end{figure}

Figure~\ref{offset_phi} shows the 2D photon angular distribution over the two spatial angles, $\theta$ and $\phi$, and the 1D projections on the azimuthal angle $\phi$, respectively. As expected, the enhancement of the radiation is found to occur in the direction of the initial offset, e.g., $\phi=\pi/4$ and $-3\pi/4$ for the case with offsets of $+0.71\sigma_{ic}$ in both the $y$- and $z$-direction as shown in Fig.~\ref{offset_phi}(a). We also notice that the off-axis injection reshapes the photon density distribution with respect to the axial angle $\theta$. In the offset direction, the photons fall in a wide range about the angle $\theta$, while in other directions about the azimuthal angle $\phi$, betatron photons are confined radially within a much narrow $\theta$ angle. For this feature, earlier researches have shown that for an electron oscillating in a plane along the propagation direction, the typical opening angle $\Theta$ of the radiation cone scales as $\Theta\sim K_\beta/\gamma$ in the trajectory plane while in the vertical direction, the typical radiation cone angle is smaller, being $\Theta\sim 1/\gamma$.
Furthermore, in Fig.~\ref{offset_phi}(b), we can find that the enhancement of the radiation along the initial offset direction leads to the reduction of photon emission in other angles of $\phi$, with respect to cases of on-axis injection. And this enhancement is also stronger for cases with higher charge. This feature may allow us to deduce the beam misalignment direction via the betatron radiation diagnostics at the plasma exit.

\section{Radiation from the electron bunch for self-modulation seeding}
\label{seedradiation}
In the first plasma cell of AWAKE Run 2 experiment, an electron bunch seeds the proton self-modulation when it runs ahead of the long proton bunch~\cite{Muggli_seeding}. The electron seeded self-modulation will offer better control on the wakefield phase and amplitude than the self-modulation grows from random noise. Furthermore, the use of a plasma density ramp can result in nearly constant wakefield after the proton bunch density modulation~\cite{PhysRevLett.125.264801, Lotov_2020}. 
Since the diagnostics of the seed electron bunch could be necessary to understand its dynamics in experiments, we investigate the suitability of betatron radiation diagnostics for the seed bunch as well.

In the following simulation, the seed electron bunch has an initial energy of 18.5 MeV, charge of 250 pC, radial size of $\sigma_{r0}=0.2\,\mathrm{mm}$ and duration of $\sigma_t=5\,\mathrm{ps}\sim1.5\,\mathrm{mm}$. The plasma density is $n_0=2\times10^{14}\,\mathrm{cm^{-3}}$. These parameters are generally similar to that used in the preliminary electron seeding experiment~\cite{seedingexp}. The plasma density step is not considered in this study. Other simulation environment settings are similar to those in the previous section. 

\begin{figure}
\centering
\includegraphics*[width=\columnwidth]{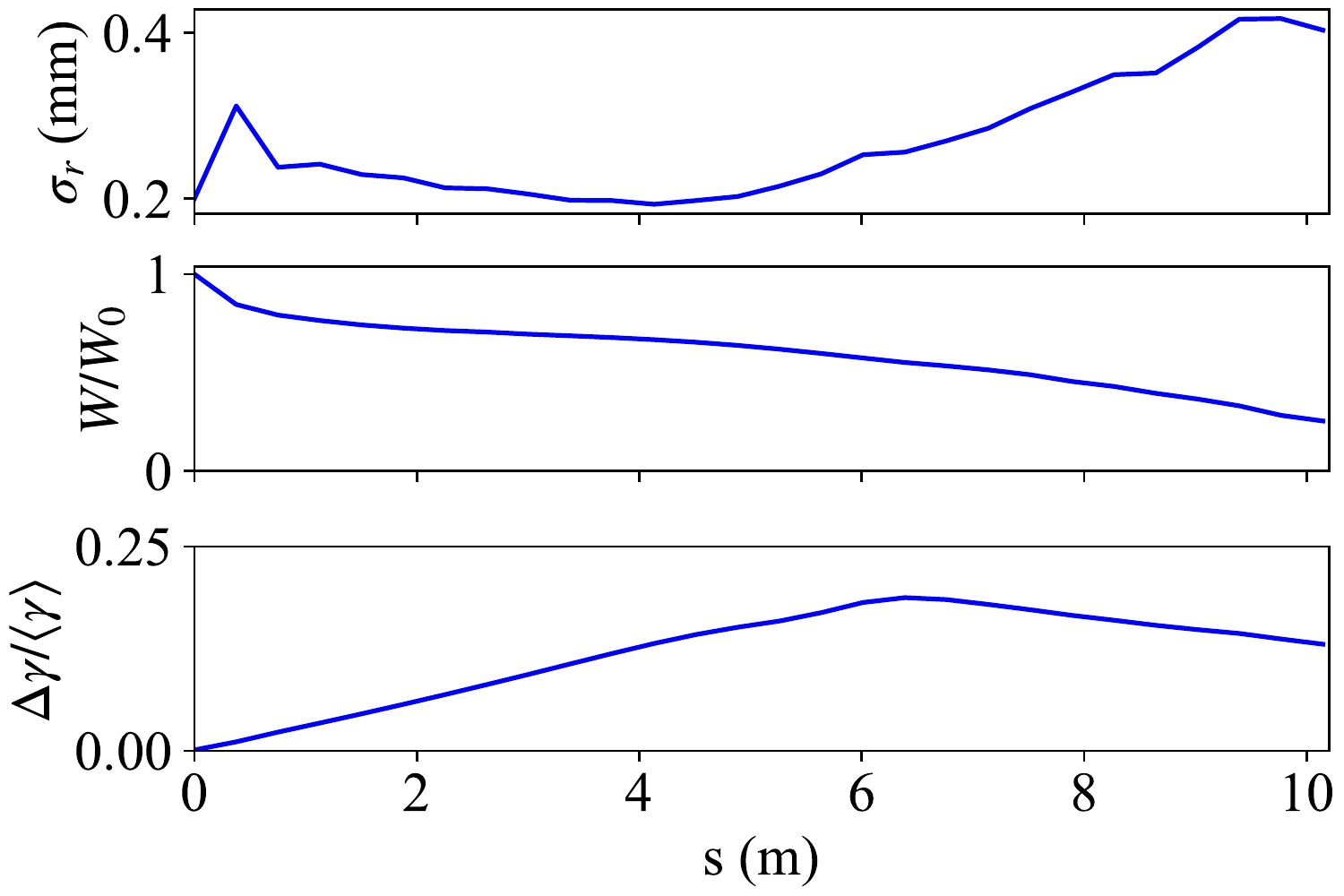}
\caption{Evolution of the seed electron bunch with respect to its propagation distance $s$ in the plasma. $\sigma_r$ is the RMS beam radius. $W=Q\left<\gamma\right>$ is the total energy stored in the seed bunch, where $Q$ and $\left<\gamma\right>$ are the charge and the mean Lorentz factor of the seed bunch. $\Delta\gamma/\left<\gamma\right>$ is the relative energy spread.
\label{seed_dynamics}
}
\end{figure}

Some of the main bunch statistics, such as the RMS radial size $\sigma_r$, total beam energy $W$ and the relative energy spread $\Delta\gamma/\left<{\gamma}\right>$, are shown in Fig.~\ref{seed_dynamics}. The fast increase of the radial RMS beam size $\sigma_r$ in the first half metre is due to the defocusing of seed bunch electrons at the head and the tail of the bunch, where the local plasma focusing force is too weak to compensate the emittance-induced defocusing effect at the beginning. This effect results in about 25\% charge leaving the simulation window without significant deceleration. After that, the remaining seed electrons get focused by the self-driven wakefield, which leads to reduction of the beam size and the increase of the seed wakefield amplitude. Since these electrons are distributed over a wide range of phases after $s=1$ m, including both the decelerating and accelerating phases, this leads to a huge increase of the beam energy spread, but no significant net deceleration effect. As the energy of electrons at the bunch head are quickly depleted in plasma after about 5 metres, they start to slip backwards into the defocusing phase, resulting in a significant bunch size expansion and charge loss. As a result of this second-stage charge loss, the remaining seed electron bunch is cooled down, as shown by the decrease of the energy spread. The overall beam energy also sees a larger decrease.

\begin{figure}
\centering
\includegraphics*[width=0.8\columnwidth]{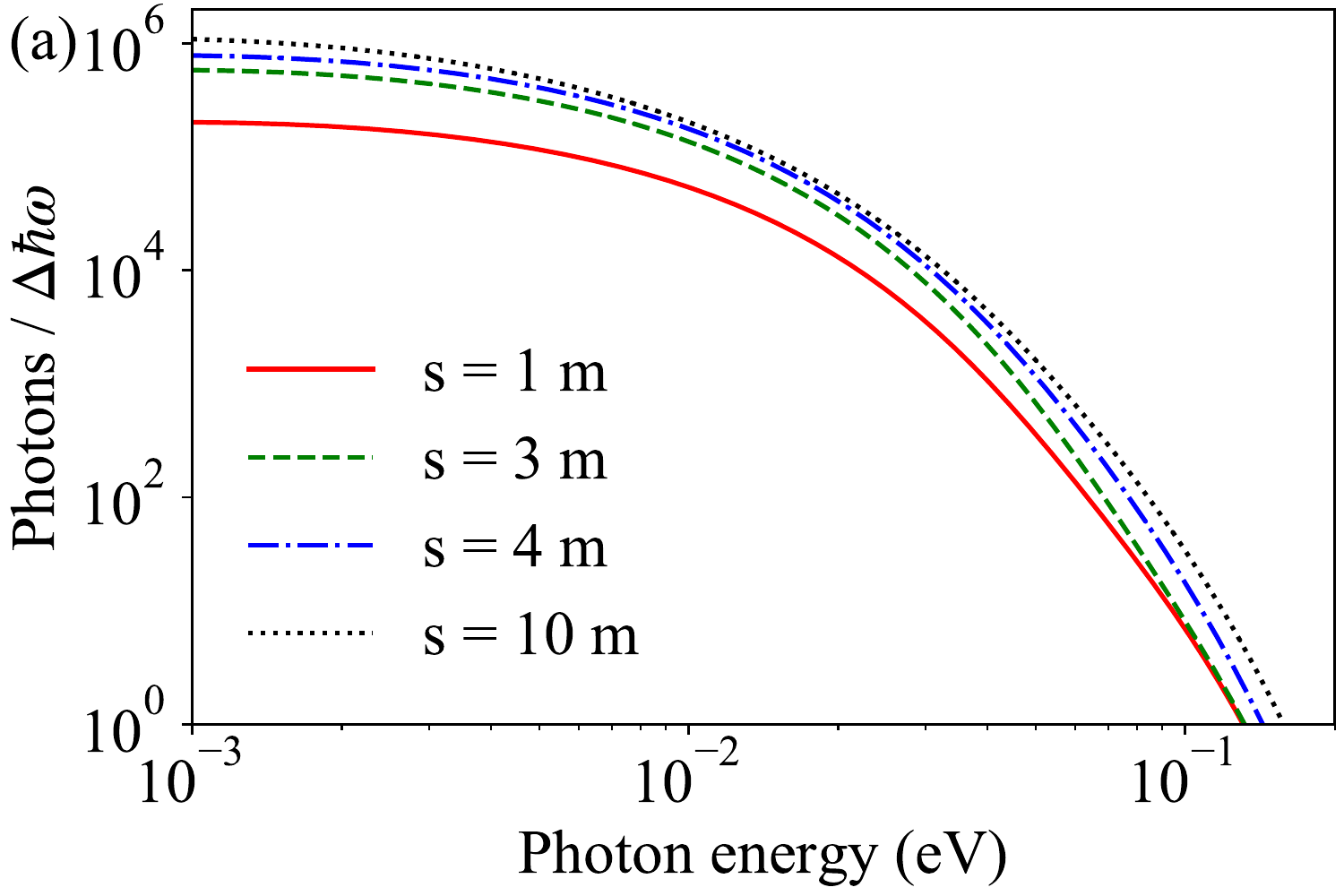}
\includegraphics*[width=0.8\columnwidth]{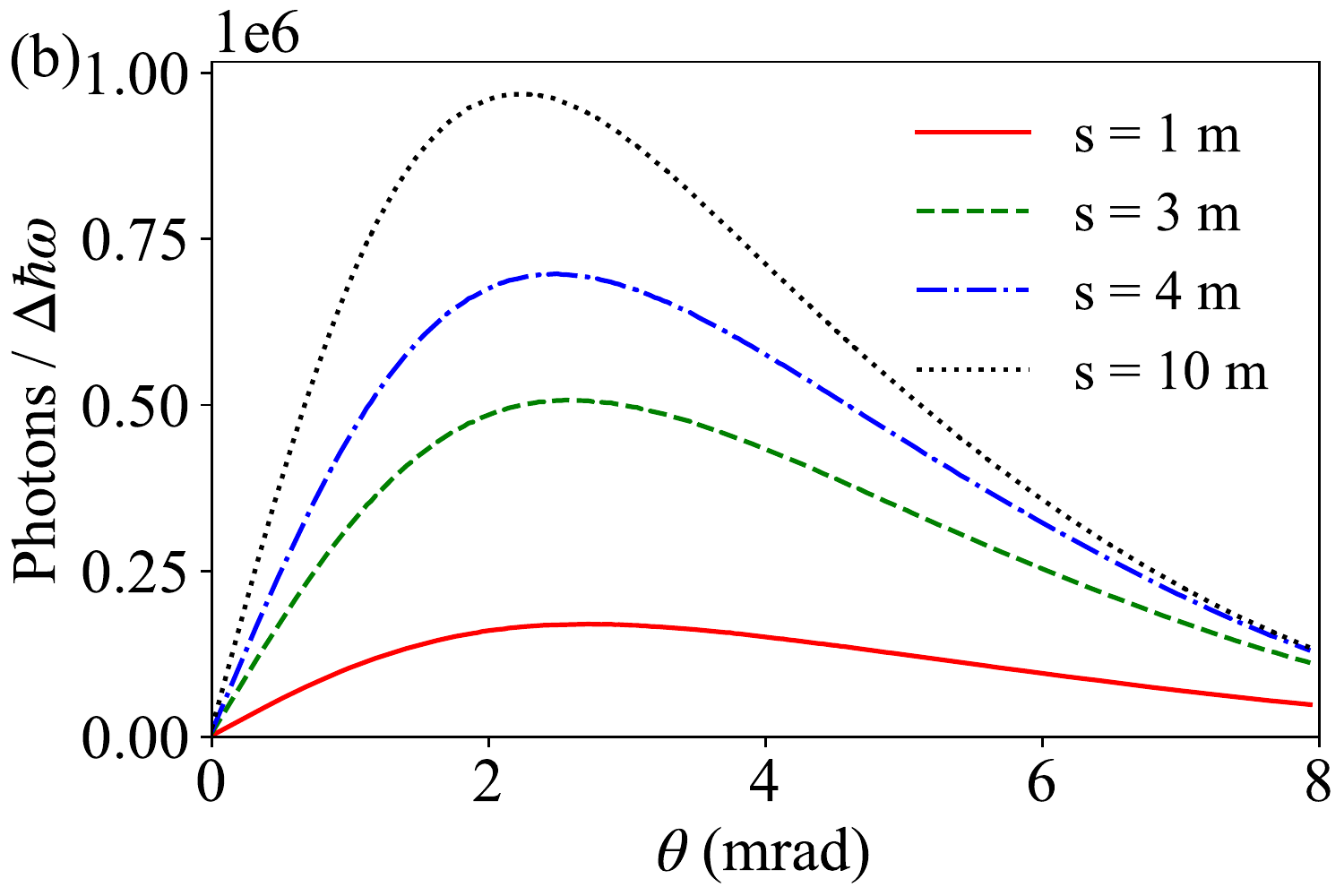}
\caption{(a) Betatron radiation spectra measured at different locations in the plasma. (b) Corresponding 1D photon angular distributions with respect to the axial angle $\theta$.
\label{seed_spectrum}
}
\end{figure}

Figure~\ref{seed_spectrum} shows the betatron photon energy spectra and the photon angular distributions of the 250 pC seed electron beam, which are measured at different locations along the self-modulation plasma cell. It can be seen that the photon energy spectra and the corresponding angular distributions don't evolve too much during the seed beam propagation. Calculation shows that the critical photon energy of the seed beam radiation is decreasing in this process, but the change is less than 10\% between $s$ = 1 m and 10 m, where the critical photon energies are 12.9 meV and 11.7 meV, respectively. Meanwhile, the maximum photon energy sees a small increase with the presence of a fraction of accelerated and defocused seed electrons. Here, the decrease of the critical photon energy is due to that the critical photon energy calculated via Eq.~\eqref{eq:critavg} is actually a renormalized average value and each of these betatron spectra shown in Fig.~\ref{seed_spectrum}(a) is the integration of all the betatron emission before the simulation diagnostic locations. So the increasing contribution of the low energy photon emission at a latter time from a larger portion of focused seed electrons that are oscillating in low amplitudes leads to the decrease of the critical photon energy of these integrated photon spectra. 
Similarly, the RMS angle $\sigma_\theta$ also slightly decreases from 4.2 mrad at $s=1$ m to 3.8 mrad at $s=10$ m. This also appears as the result of the increasing of betatron emission from low-amplitude seed electron oscillations.

As the measured critical photon energy and the RMS angle of the angular photon distribution don't evolve too much, we might be able to use the betatron photon spectrum measured at the exit of the self-modulation stage to estimate the radial size of the seed bunch at an earlier time. 

\section{Discussion}
\label{discussection}
The correlation between betatron radiation and the electron dynamics has shown that we might be able to use betatron radiation to indirectly measure the evolution of the beam size or even beam profile inside the plasma tube. However, as also shown by this work and the previous study~\cite{WILLIAMSON2020164076}, there are several difficulties in the study and the application of betatron diagnostics for proton wakefield acceleration experiments. The main problem from the physical side is the acceleration-leaded witness beam evolution, which then changes the characteristics of the betatron radiation emitted at each moment along the acceleration path. When these betatron photons emitted from different electrons and at different times are accumulated on the screen of the spectrometer, the integrated radiation spectrum can no longer be simply characterized by the single electron betatron radiation theory. Moreover, the majority of the witness electrons with baseline parameters are oscillating in the quasi-undulator regime ($K_\beta\sim1$) during the acceleration. This again makes the asymptotic expression of the single electron radiation less accurate for fitting with the simulation data. Nonetheless, if a high time resolution spectrometer to resolve the emission at different times is available, the diagnostics of witness bunch betatron emission can still qualitatively reveal the beam envelope evolution inside the plasma.

As the proton bunch or bunch train coexists with electron bunch in both stage of the Run 2 experiment setup, it is also interesting to look at the radiation from the proton bunches. 
The main difficulty for simulation study of proton bunch radiation is the huge simulation cost due to relatively large scale of the problem. 
Instead, we can give an estimation of the typical betatron photon energy radiated by these protons. For a 400 GeV proton oscillating from an initial radial position of one $\sigma_{rp}$ = 0.2 mm in the plasma ion column, where the plasma density is $2\times10^{14}\,\mathrm{cm^{-3}}$, the critical photon energy can be calculated via Eq.~\eqref{Eq. criteng}. As the proton-driven wakefield is in the quasi-linear regime, where the plasma focusing force is weaker than in the pure ion column thus the actual normalized betatron oscillation amplitude $K_\beta$ is also lower, this calculation gives an upper limit of the betatron photon energy as $\hbar\omega_c=0.089$ eV. Since this estimated critical photon energy of the radiation emitted by proton is almost in the same range as that of the seed electron bunch radiation, it may therefore be difficult to distinguish the betatron radiation from two kinds of particles. However, for the betatron radiation diagnostics at the acceleration stage where the electron radiation energy, typically in VUV ($>10\,\mathrm{eV}$) to X-ray ($\sim\mathrm{keV}$) range, is much higher than that of protons (being 0.313 eV in the high density plasma, i.e., $n_0=7\times10^{14}\,\mathrm{cm^{-3}}$), this would not be a problem. 

Another possible source of interference for betatron radiation diagnostics could be the low energy photons emitted by oscillating plasma background electrons~\cite{doi:10.1063/1.1624605}. However, since this part of radiation is more possible to originate from the oscillations of low energy plasma electrons in the quasilinear wakefields, and the acceleration of plasma electrons due to the self-injection effect in the ultra nonlinear regime~\cite{PhysRevLett.103.175003} is unlikely to happen in the Run 2 scheme. Therefore, the interference from background plasma electron radiation is expected to be negligible.

\section{Conclusion}
In this article, we studied the betatron radiation from the electron bunches in the AWAKE Run 2. 
The simulation results show the betatron radiation can effectively reveal the evolution of the witness electron beam properties, such as the RMS radial size and average beam energy. In addition, betatron radiation diagnostics for the acceleration stage of AWAKE Run 2 can also identify the radial bunch size mismatch and offset of the witness electron bunch at the injection point. This may work as a useful feedback for the injection system. Additionally, we examined the possibility of using the radiation from the seed electron bunch in the self-modulation stage for betatron diagnostics. However, the presence of the proton betatron radiation in the same frequency range makes it difficult to do so. Nevertheless, our work provides a further understanding of the betatron radiation properties in the AWAKE Run 2 and contributes to the study of betatron radiation diagnostics for future proton-driven wakefield accelerators.

\begin{acknowledgments}
The authors would like to acknowledge the support from the Cockcroft Institute Core Grant and the STFC AWAKE Run 2 grant ST/T001917/1. Computing resources are provided by the SCARF HPC of STFC and the CERN HPC services. The authors would also like to thank the members of the AWAKE Collaboration for helpful discussions.

\end{acknowledgments}

\bibliography{Refs.bib}

\end{document}